\documentclass[a4paper,12pt]{article}

\usepackage{ifpdf}

\newif\ifpdf
\ifx\pdfoutput\undefined
  \pdffalse
\else
  \pdfoutput=1
  \pdftrue
\fi

\RequirePackage{xspace} %
\RequirePackage{subfigure} %
\RequirePackage[centertags]{amsmath} %
\RequirePackage{amssymb}
\RequirePackage{wrapfig} %
\RequirePackage{calc} %
\RequirePackage{ifthen}
\RequirePackage{tabularx} %
\RequirePackage{flafter} %
\RequirePackage{fancyhdr} %

\ifpdf
  \RequirePackage[pdftex]{color}%
  \RequirePackage{colortbl}%
  \RequirePackage{array}%
  \RequirePackage[pdftex]{graphicx}

  \RequirePackage[ pdftex, plainpages = false, pdfpagelabels,
                 pdfpagelayout = useoutlines,
                 bookmarks,
                 breaklinks = true,
                 linktocpage,
                 pagebackref,                      
                 colorlinks = true,
                 linkcolor = blue,
                 urlcolor  = blue,
                 citecolor = blue,
                 anchorcolor = blue,
                 hyperindex = true,
                 hyperfigures
                 ]{hyperref}

\else
  \RequirePackage{color}
  \RequirePackage{colortbl}
   \RequirePackage{array}
  \RequirePackage[dvips]{graphicx}
  \RequirePackage{hyperref}
  \usepackage{rotating}
\fi

\usepackage{makeidx} 
\usepackage{setspace} 
\usepackage{rotating} 
\usepackage{ecltree}
\usepackage{epic}
\usepackage{supertabular}  
\usepackage{color}
\usepackage{exscale}
\usepackage{fontenc}
\usepackage{ifthen}
\usepackage{latexsym}
\usepackage{makeidx}
\usepackage{syntonly}
\usepackage{inputenc}
\usepackage{graphicx}
\usepackage{setspace}
\usepackage{caption2}
\usepackage[english]{babel}
\usepackage[square, comma,numbers,sort&compress]{natbib}
\usepackage{hypernat}
\usepackage{boxedminipage}
\usepackage{framed}
\usepackage{longtable}
\usepackage[all]{hypcap}    
\usepackage{algorithm2e}
\usepackage{algorithmic}
\usepackage{lscape}
\usepackage{pdflscape}

\newcommand{\note}[1]{\marginpar[left]{\singlespace \tiny #1}}

\renewcommand{\sectionmark}[1]%
      {\markright{\thesection\ #1}} 

\renewcommand{\note}[1]{}

\doublespace 

\title
 {
\vspace*{5.0cm} \Large{\bf  One-Dimensional Navier-Stokes Finite Element Flow Model} \vspace*{3.0cm} \\
\Large{\bf Taha Sochi\footnote{Imaging Sciences \& Biomedical Engineering, King's College London,
St Thomas' Hospital, London, SE1 7EH, UK. Email: taha.sochi@kcl.ac.uk.}} \vspace*{3.0cm} \\
\large{\bf  Technical Report}
}

\makeindex

\begin{document}

\maketitle %
\pagenumbering{arabic}

\newpage
\phantomsection \addcontentsline{toc}{section}{Contents} %
\tableofcontents

%
\phantomsection \addcontentsline{toc}{section}{List of Tables} %
\listoftables

\newpage
\phantomsection \addcontentsline{toc}{section}{Abstract} \noindent
{\noindent \LARGE \bf Abstract} \vspace{0.5cm}\\
\noindent
This technical report documents the theoretical, computational, and practical aspects of the
one-dimensional Navier-Stokes finite element flow model. The document is particularly useful to
those who are interested in implementing, validating and utilizing this relatively-simple and
widely-used model.

\vspace{0.5cm}


\pagestyle{headings} %
\addtolength{\headheight}{+1.6pt}
\lhead[{Chapter \thechapter \thepage}]%
      {{\bfseries\rightmark}}
\rhead[{\bfseries\leftmark}]%
     {{\bfseries\thepage}} 
\headsep = 1.0cm               

\newpage
\section{Introduction}

The one-dimensional (1D) Navier-Stokes flow model in its analytic formulation and numeric
implementation is widely used for calculating and simulating the flow of Newtonian fluids in large
vessels and in interconnected networks of such vessels \cite{OlufsenPKPNL2000, SherwinFPP2003,
FormaggiaLTV2006, JanelaMS2010, SochiNavier2013}. In particular, the model is commonly used by
bioengineers to analyze blood flow in the arteries and veins. This model can be easily implemented
using a numeric meshing technique, such as finite element method, to provide a computational
framework for flow simulation in large tubes. The model can also be coupled with a pressure-area
constitutive relation and hence be extended to elastic vessels and networks of elastic vessels.
Despite its simplicity, the model is reliable within its validity domain and hence it can provide
an attractive alternative to the more complex and costly multi-dimensional flow models in some
cases of flow in regular geometries with obvious symmetry.

The roots of the 1D flow model may be traced back to the days of Euler who apparently laid down its
mathematical foundations. In the recent years, the 1D model became increasingly popular, especially
in the hemodynamics modeling. This is manifested by the fact that several researchers
\cite{FormaggiaGNQ2001, SmithPH2002, RuanCZC2003, SherwinFPP2003, UrquizaBLVF2003, PontrelliR2003,
MilisicQ2004, FernandezMQ2005, FormaggiaLTV2006, FormaggiaMN2006, AlastrueyMPDPS2007,
AlastrueyPPS2008, LeeS2008, PasseriniLFQV2009, Papadakis2009, JanelaMS2010} have used this model
recently in their modeling and simulation work.

The `1D' label attached to this model stems from the fact that the $\theta$ and $r$ dependencies of
a cylindrically-coordinated vessel are neglected due to the axi-symmetric flow assumption and the
simplified consideration of the flow profile within a lumped parameter called the momentum
correction factor. Therefore, the only dependency that is explicitly accounted for is the
dependency in the flow direction, $x$.

The biggest advantages of the 1D model are the relative ease of implementation, and comparative low
computational cost in execution. Therefore, the use of full multi-dimensional flow modeling,
assuming its viability within the available computational resources, is justified only when the 1D
model fails to capture the essential physical picture of the flow system. However, there are
several limitations and disadvantages of the 1D model that restrict its use. These limitations
include, among other things, the Newtonian assumption, simplified flow geometry and the
one-dimensional nature.

\section{Theoretical Background}

The widely-used one-dimensional form of the Navier-Stokes equations to describe the flow in a
vessel; assuming laminar, incompressible, axi-symmetric, Newtonian, fully-developed flow with
negligible gravitational body forces; is given by the following continuity and momentum balance
relations with suitable boundary conditions

\begin{eqnarray}
\frac{\partial A}{\partial t}+\frac{\partial Q}{\partial x}&=&0\,\,\,\,\,\,\,\,\,\,\,\,\,
t\ge0,\,\,\, x\in[0,L]     \label{ConEq1} \\
\frac{\partial Q}{\partial t}+\frac{\partial}{\partial x}\left(\frac{\alpha
Q^{2}}{A}\right)+\frac{A}{\rho}\frac{\partial p}{\partial
x}+\kappa\frac{Q}{A}&=&0\,\,\,\,\,\,\,\,\,\,\,\,\, t\ge0,\,\,\, x\in[0,L]     \label{MomEq1}
\end{eqnarray}
In these equations, $A$ is the vessel cross sectional area, $t$ is the time, $Q$ is the volumetric
flow rate, $x$ is the axial coordinate along the vessel, $L$ is the length of the vessel, $\alpha$
($=\frac{\int u^{2}dA}{A\overline{u}^{2}}$ with $u$ and $\overline{u}$ being the fluid local and
mean axial speed respectively) is the momentum flux correction factor, $\rho$ is the fluid mass
density, $p$ is the local pressure, and $\kappa$ is a viscosity friction coefficient which is
usually given by $\kappa = 2\pi\alpha\nu/(\alpha-1)$ with $\nu$ being the fluid kinematic viscosity
defined as the ratio of the dynamic viscosity $\mu$ to the mass density. These equations supported
by appropriate compatibility and matching conditions are used to describe the 1D flow in a branched
network of vessels. The equations, being two in three variables, $Q$ $A$ and $p$, are normally
coupled with the following pressure-area relation in a distensible vessel to close the system and
obtain a solution

\begin{equation}\label{PAEq}
p = p_o + f(A)
\end{equation}
In this relation, $p$ and $p_o$ are the local and reference pressure respectively, and $f(A)$ is a
function of area which may be modeled by the following relation

\begin{equation}\label{fAEq}
f(A)=\frac{\beta}{A_{o}}\left(\sqrt{A}-\sqrt{A_{o}}\right)
\end{equation}
where
\begin{equation}
\beta=\frac{\sqrt{\pi}h_{o}E}{1-\varsigma^{2}}
\end{equation}
In these equations, $A_o$ and $h_o$ are respectively the vessel cross sectional area and wall
thickness at reference pressure $p_o$, while $E$ and $\varsigma$ are the Young's elastic modulus
and Poisson's ratio of the vessel wall. Similar variants of this 1D flow model formulation can also
be found in the literature (see for example \cite{CarloNPT, SmithPH2002, PontrelliR2003,
Papadakis2009}).

The continuity and momentum equations are usually casted in matrix form \cite{SherwinFPP2003,
MilisicQ2004, FernandezMQ2005, LeeS2008} which is more appropriate for numerical manipulation and
discretization. In matrix form these equations are given by

\begin{equation}\label{MatrixEq1}
\frac{\partial\mathbf{U}}{\partial t}+\frac{\partial\mathbf{F}}{\partial x}+\mathbf{B}=\mathbf{0}
\end{equation}
where

\begin{equation}\label{MatrixEq2}
\mathbf{U}=\left[\begin{array}{c}
A\\
Q\end{array}\right]\,\,,\,\,\,\,\,\,\,\,\,\,\,\,\,\,\,\mathbf{F}=\left[\begin{array}{c}
Q\\
\frac{\alpha
Q^{2}}{A}+\int_{A}\frac{a}{\rho}\frac{df}{da}da\end{array}\right]=\left[\begin{array}{c}
Q\\
\frac{\alpha Q^{2}}{A}+\frac{\beta}{3\rho A_o}A^{3/2}\end{array}\right]
\end{equation}
and

\begin{equation}\label{MatrixEq3}
\mathbf{B}=\left[\begin{array}{c}
0\\
\kappa\frac{Q}{A}\end{array}\right]
\end{equation}

It should be remarked that the second term of the second row of the $\mathbf{F}$ matrix can be
obtained from the third term of the original momentum equation as follow

\begin{equation}\nonumber
\frac{A}{\rho}\frac{\partial p}{\partial x}=\frac{A}{\rho}\frac{\partial f}{\partial
x}=\frac{A}{\rho}\frac{\partial f}{\partial A}\frac{\partial A}{\partial
x}=\frac{\partial}{\partial x}\int_{x'}\frac{A}{\rho}\frac{\partial f}{\partial A}\frac{\partial
A}{\partial x}\partial x
\end{equation}

\begin{equation}\label{thirdTermF}
=\frac{\partial}{\partial x}\int_{A'}\frac{A}{\rho}\frac{\partial f}{\partial A}\partial
A=\frac{\partial}{\partial x}\int_{A'}\frac{A}{\rho}\frac{df}{dA}dA=\frac{\partial}{\partial
x}\left(\frac{\beta}{3\rho A_{o}}A^{3/2}\right)
\end{equation}

\section{Weak Form of 1D Flow Equations}

On multiplying Equation \ref{MatrixEq1} by weight functions and integrating over the solution
domain, $x$, the following is obtained

\begin{equation}\label{WeakFormEq1}
\int_{\Omega}\frac{\partial\mathbf{U}}{\partial
t}\cdot\boldsymbol{\omega}dx+\int_{\Omega}\frac{\partial\mathbf{F}}{\partial
x}\cdot\boldsymbol{\omega}dx+\int_{\Omega}\mathbf{B}\cdot\boldsymbol{\omega}dx=\mathbf{0}
\end{equation}
where $\Omega$ is the solution domain, and $\boldsymbol{\omega}$ is a vector of arbitrary test
functions. On integrating the second term of Equation \ref{WeakFormEq1} by parts, the following
weak form of the preceding 1D flow system is obtained

\begin{equation}\label{WeakFormEq2}
\int_{\Omega}\frac{\partial\mathbf{U}}{\partial
t}\cdot\boldsymbol{\omega}dx-\int_{\Omega}\mathbf{F}\cdot\frac{d\boldsymbol{\omega}}{dx}dx+\int_{\Omega}\mathbf{B}\cdot\boldsymbol{\omega}dx+[\mathbf{F}\cdot\boldsymbol{\omega}]_{\partial\Omega}=\mathbf{0}
\end{equation}
where $\partial \Omega$ is the boundary of the solution domain. This weak formulation, coupled with
suitable boundary conditions, can be used as a basis for finite element implementation in
conjunction with an iterative scheme such as Newton-Raphson method.

\section{Finite Element Solution}

There are two major cases to be considered in the finite element solution of the stated flow
problem: single vessel and branched network where each one of these cases can be time-independent
or time-dependent. These four cases are outlined in the following three subsections.

\subsection{Single Vessel Time-Independent Flow}

The single vessel time-independent model is based on dropping the time term in the continuity and
momentum governing equations to obtain a steady-state solution. This should be coupled with
pertinent boundary and compatibility conditions at the vessel inlet and outlet. The details are
given in the following.

In discretized form, Equation \ref{WeakFormEq2} without the time term can be written for each node
$N_{i}(A_{i},Q_{i})$ as

\begin{equation}
\mathbf{R}_{i}=\left[\begin{array}{c}
f_{i}\\
g_{i}\end{array}\right]=\left[\begin{array}{c}
0\\
0\end{array}\right]\end{equation}
where $\mathbf{R}$ is a vector of the weak form of the residuals and

\begin{equation}
f_{i}=\sum_{q=1}^{N_{q}}\left[-w_{q}Q(\zeta_{q})\frac{\partial
x}{\partial\zeta}(\zeta_{q})\frac{d\omega_{A_{i}}}{d\zeta}(\zeta_{q})\frac{d\zeta}{dx}(\zeta_{q})\right]+Q(\partial\Omega)\omega_{A_{i}}(\partial\Omega)
\end{equation}

and

\begin{eqnarray}
g_{i} & = & \sum_{q=1}^{N_{q}}w_{q}\frac{\partial x}{\partial\zeta}(\zeta_{q})\left[-\left(\frac{\alpha Q^{2}(\zeta_{q})}{A(\zeta_{q})}+\frac{\beta}{3\rho A_o}A^{3/2}(\zeta_{q})\right)\frac{d\omega_{Q_{i}}}{d\zeta}(\zeta_{q})\frac{d\zeta}{dx}(\zeta_{q})+\kappa\frac{Q(\zeta_{q})}{A(\zeta_{q})}\omega_{Q_{i}}(\zeta_{q})\right] \nonumber \\
 &  & +\left(\frac{\alpha Q^{2}(\partial\Omega)}{A(\partial\Omega)}+\frac{\beta}{3\rho A_o}A^{3/2}(\partial\Omega)\right)\omega_{Q_{i}}(\partial\Omega)
\end{eqnarray}
where $q$ is an index for the $N_{q}$ quadrature points, $\zeta$ is the quadrature point coordinate
and

\begin{equation}
A(\zeta_{q})=\sum_{i}^{n}A_{c_{i}}\psi_{A_{i}}(\zeta_{q})\,\,\,\,\,\,\,\,\,\,\,\,\, \&
\,\,\,\,\,\,\,\,\,\,\,\,\, Q(\zeta_{q})=\sum_{i}^{n}Q_{c_{i}}\psi_{Q_{i}}(\zeta_{q})
\end{equation}
with $n$ being the number of nodes in a standard element. Because of the non-linear nature of the
problem, an iteration scheme, such as  Newton-Raphson, can be utilized to construct and solve this
system of equations based on the residual. The essence of this process is to solve the following
equation iteratively and update the solution until a convergence criterion based on reaching a
predefined error tolerance is satisfied

\begin{equation}\label{JDUR}
\mathbf{J} \, \Delta\mathbf{U}=-\mathbf{R}
\end{equation}

In this equation, $\mathbf{J}$ is the jacobian matrix, $\Delta\mathbf{U}$ is the perturbation
vector, and $\mathbf{R}$ is the weak form of the residual vector. For a vessel with $n$ nodes, the
Jacobian matrix, which is of size $2n\times2n$, is given by

\begin{equation}
\mathbf{J}=\left[\begin{array}{ccccc}
\frac{\partial f_{1}}{\partial A_{1}} & \frac{\partial f_{1}}{\partial Q_{1}} & \cdots & \frac{\partial f_{1}}{\partial A_{n}} & \frac{\partial f_{1}}{\partial Q_{n}}\\
\frac{\partial g_{1}}{\partial A_{1}} & \frac{\partial g_{1}}{\partial Q_{1}} & \cdots & \frac{\partial g_{1}}{\partial A_{n}} & \frac{\partial g_{1}}{\partial Q_{n}}\\
\vdots & \vdots & \ddots & \vdots & \vdots\\
\frac{\partial f_{n}}{\partial A_{1}} & \frac{\partial f_{n}}{\partial Q_{1}} & \cdots & \frac{\partial f_{n}}{\partial A_{n}} & \frac{\partial f_{n}}{\partial Q_{n}}\\
\frac{\partial g_{n}}{\partial A_{1}} & \frac{\partial g_{n}}{\partial Q_{1}} & \cdots &
\frac{\partial g_{n}}{\partial A_{n}} & \frac{\partial g_{n}}{\partial
Q_{n}}\end{array}\right]
\end{equation}
where the subscripts stand for the node indices, while the vector of unknowns, which is of size
$2n$, is given by

\begin{equation}
\mathbf{U}=\left[\begin{array}{c}
A_{1}\\
Q_{1}\\
\vdots\\
A_{n}\\
Q_{n}\end{array}\right]
\end{equation}

In the finite element implementation, the Jacobian matrix is usually evaluated numerically by
finite differencing, i.e.

\begin{equation}\label{jacobEq}
\mathbf{J}\simeq\left[\begin{array}{ccccc}
\frac{\Delta f_{1}}{\Delta A_{1}} & \frac{\Delta f_{1}}{\Delta Q_{1}} & \cdots & \frac{\Delta f_{1}}{\Delta A_{n}} & \frac{\Delta f_{1}}{\Delta Q_{n}}\\
\frac{\Delta g_{1}}{\Delta A_{1}} & \frac{\Delta g_{1}}{\Delta Q_{1}} & \cdots & \frac{\Delta g_{1}}{\Delta A_{n}} & \frac{\Delta g_{1}}{\Delta Q_{n}}\\
\vdots & \vdots & \ddots & \vdots & \vdots\\
\frac{\Delta f_{n}}{\Delta A_{1}} & \frac{\Delta f_{n}}{\Delta Q_{1}} & \cdots & \frac{\Delta f_{n}}{\Delta A_{n}} & \frac{\Delta f_{n}}{\Delta Q_{n}}\\
\frac{\Delta g_{n}}{\Delta A_{1}} & \frac{\partial g_{n}}{\Delta Q_{1}} & \cdots & \frac{\Delta
g_{n}}{\Delta A_{n}} & \frac{\Delta g_{n}}{\Delta Q_{n}}\end{array}\right]
\end{equation}

The procedure to obtain a solution is summarized in the following scheme

\begin{enumerate}

\item
Start with initial values for $A_i$ and $Q_i$ in the $\mathbf{U}$ vector.

\item
The system given by Equation \ref{JDUR} is constructed where the weak form of the residual vector
$\mathbf{R}$ may be calculated in each iteration $l$ ($=0,1,\ldots,M$) as

\begin{equation}
\mathbf{R}_{l}=\left[\begin{array}{c}
f_{1}(\mathbf{U}_{l})\\
g_{1}(\mathbf{U}_{l})\\
\vdots\\
f_{n}(\mathbf{U}_{l})\\
g_{n}(\mathbf{U}_{l})\end{array}\right]
\end{equation}

\item
The jacobian matrix is calculated from Equation \ref{jacobEq}.

\item
System \ref{JDUR} is solved for $\Delta\mathbf{U}$, i.e.

\begin{equation}
\Delta\mathbf{U}=-\mathbf{J}^{-1}\mathbf{R}
\end{equation}

\item
$\mathbf{U}$ is updated to obtain a new $\mathbf{U}$ for the next iteration, that is

\begin{equation}
\mathbf{U}_{l+1}=\mathbf{U}_{l}+\Delta\mathbf{U}
\end{equation}

\item The norm of the residual vector is calculated from

\begin{equation}
\mathfrak{N}=\frac{\sqrt{\epsilon_{1}^{2}+\epsilon_{2}^{2}+\cdots+\epsilon_{N}^{2}}}{N}
\end{equation}
where $\epsilon_{i}$ is the $i$th entry of the residual vector and $N$ ($=2n$) is the size of the
residual vector.

\item
This cycle is repeated until the norm is less than a predefined error tolerance (e.g. $10^{-8}$) or
a certain number of cycles is reached without convergence. In the last case, the operation will be
aborted due to failure and may be resumed with improved finite element parameters.

\end{enumerate}

\vspace{0.5cm}

With regard to the boundary conditions (BC), two types of Dirichlet conditions can be applied:
pressure and volumetric flow rate, that is
\begin{equation}\label{TIBC}
A-A_{BC}=0 \hspace{0.5cm} ({\rm for\,\,area\,\,BC}) \hspace{0.5cm} \& \hspace{0.5cm} Q-Q_{BC}=0
\hspace{0.5cm} ({\rm for\,\,flow\,\,BC})
\end{equation}

These conditions are imposed by replacing the residual function of one of the governing equations
(the continuity equation in our model) for the boundary nodes with one of these constraints.

Imposing the boundary conditions as constraints in one of the two governing equations is associated
with imposing compatibility conditions, arising from projecting the differential equations in the
direction of the outgoing characteristic variables \cite{FormaggiaLQ2003}, at the inlet and outlet
by replacing the residual function contributed by the other governing equation with these
conditions. The compatibility conditions are given by

\begin{equation}\label{BCMEq}
l_{1,2}^{T}\left(\mathbf{H}\frac{\partial\mathbf{U}}{\partial x}+\mathbf{B}\right)=0
\end{equation}
where $\mathbf{H}$ is the matrix of partial derivative of $\mathbf{F}$ with respect to
$\mathbf{U}$, while the transposed left eigenvectors of $\mathbf{H}$ are given by

\begin{equation}
l_{1,2}^{T}=\left[\begin{array}{cc}
-\alpha\frac{Q}{A}\pm\sqrt{\frac{Q^{2}}{A^{2}}\left(\alpha^{2}-\alpha\right)+\frac{A}{\rho}\frac{df}{dA}}
& 1\end{array}\right]
\end{equation}
that is

\begin{equation}
\mathbf{H}\frac{\partial\mathbf{U}}{\partial x}+\mathbf{B}=\left[\begin{array}{cc}
0 & 1\\
-\alpha\frac{Q^{2}}{A^{2}}+\frac{\beta}{2\rho A_{o}}A^{1/2} &
2\alpha\frac{Q}{A}\end{array}\right]\left[\begin{array}{c}
\frac{\partial A}{\partial x}\\
\frac{\partial Q}{\partial x}\end{array}\right]+\left[\begin{array}{c}
0\\
\kappa\frac{Q}{A}\end{array}\right]
\end{equation}
i.e.

\begin{equation}
\mathbf{H}\frac{\partial\mathbf{U}}{\partial x}+\mathbf{B}=\left[\begin{array}{c}
\frac{\partial Q}{\partial x}\\
\left(-\alpha\frac{Q^{2}}{A^{2}}+\frac{\beta}{2\rho A_{o}}A^{1/2}\right)\frac{\partial A}{\partial
x}+\left(2\alpha\frac{\partial Q}{\partial x}+\kappa\right)\frac{Q}{A}\end{array}\right]
\end{equation}

Hence, Equation \ref{BCMEq} reduces to

\begin{equation}
\left[\begin{array}{cc}
-\alpha\frac{Q}{A}\pm\sqrt{\frac{Q^{2}}{A^{2}}\left(\alpha^{2}-\alpha\right)+\frac{A}{\rho}\frac{df}{dA}}
& 1\end{array}\right]\left[\begin{array}{c}
\frac{\partial Q}{\partial x}\\
\left(-\alpha\frac{Q^{2}}{A^{2}}+\frac{\beta}{2\rho A_{o}}A^{1/2}\right)\frac{\partial A}{\partial
x}+\left(2\alpha\frac{\partial Q}{\partial
x}+\kappa\right)\frac{Q}{A}\end{array}\right]=0
\end{equation}
that is

\begin{equation}\label{TICC}
\left(-\alpha\frac{Q}{A}\pm\sqrt{\frac{Q^{2}}{A^{2}}\left(\alpha^{2}-\alpha\right)+\frac{A}{\rho}\frac{df}{dA}}\right)\frac{\partial
Q}{\partial x}+\left(-\alpha\frac{Q^{2}}{A^{2}}+\frac{\beta}{2\rho
A_{o}}A^{1/2}\right)\frac{\partial A}{\partial x}+\left(2\alpha\frac{\partial Q}{\partial
x}+\kappa\right)\frac{Q}{A}=0
\end{equation}

In the last relation, the minus sign is used for the inflow boundary while the plus sign for the
outflow boundary. The compatibility conditions, given by Equation \ref{TICC}, replace the momentum
residual at the boundary nodes.

\subsection{Single Vessel Time-Dependent Flow}

The aforementioned time-independent formulation can be extended to describe transient states by
including the time terms in the residual equations in association with a numerical time-stepping
method such as forward Euler, or backward Euler or central difference. The time-dependent residual
will then be given (in one of the aforementioned schemes) by

\begin{equation}
\mathbf{R}_{TD}^{t+\Delta t}=\int_{\Omega}\frac{\mathbf{U}^{t+\Delta t}-\mathbf{U}^{t}}{\Delta
t}\cdot\boldsymbol{\omega}dx+\mathbf{R}_{TI}^{t+\Delta t}=\mathbf{0}
\end{equation}
where $\mathbf{R}$ is the weak form of the residual, $TD$ stands for time-dependent and $TI$ for
time-independent. The time-dependent jacobian follows

\begin{equation}
\mathbf{J}_{TD}^{t+\Delta t}=\frac{\partial\mathbf{R}_{TD}^{t+\Delta
t}}{\partial\mathbf{U}^{t+\Delta t}}
\end{equation}

Again, we have

\begin{equation}
\Delta\mathbf{U}=-\mathbf{J}^{-1}\mathbf{R}
\end{equation}
and

\begin{equation}
\mathbf{U}_{l+1}=\mathbf{U}_{l}+\Delta\mathbf{U}
\end{equation}
where the symbols represent time-dependent quantities and $l$ represents Newton iterations.

With regard to the boundary nodes, a steady-state or time-dependent boundary conditions could be
applied depending on the physical situation while a time-dependent compatibility conditions should
be employed by adding a time term to the time-independent compatibility condition, that is
\begin{equation}
C_{TD}=l_{1,2}^{T}\frac{\partial\mathbf{U}}{\partial t}+C_{TI}=0
\end{equation}
where $C_{TI}$ is the time-independent compatibility condition as given by Equation \ref{TICC},
while $C_{TD}$ is the time-dependent compatibility condition, that is

\begin{equation}
C_{TD}=\left[\begin{array}{cc}
-\alpha\frac{Q}{A}\pm\sqrt{\frac{Q^{2}}{A^{2}}\left(\alpha^{2}-\alpha\right)+\frac{A}{\rho}\frac{df}{dA}}
& 1\end{array}\right]\left[\begin{array}{c}
\frac{\partial A}{\partial t}\\
\frac{\partial Q}{\partial t}\end{array}\right]+C_{TI}=0
\end{equation}
i.e.

\begin{equation}\label{TDCompatibilityEq}
C_{TD}=\left(-\alpha\frac{Q}{A}\pm\sqrt{\frac{Q^{2}}{A^{2}}\left(\alpha^{2}-\alpha\right)+\frac{A}{\rho}\frac{df}{dA}}\right)\frac{\partial
A}{\partial t}+\frac{\partial Q}{\partial t}+C_{TI}=0
\end{equation}
where the time derivatives can be evaluated by finite difference, e.g.

\begin{equation}
\frac{\partial A}{\partial t}\simeq\frac{A^{t+\Delta t}-A^{t}}{\Delta
t}\,\,\,\,\,\,\,\,\,\,\,\,\,\,\,\&\,\,\,\,\,\,\,\,\,\,\,\,\,\,\,\,\frac{\partial Q}{\partial
t}\simeq\frac{Q^{t+\Delta t}-Q^{t}}{\Delta t}
\end{equation}
A sign convention similar to that outlined previously should be followed. An algorithmic code of
the time-dependent module is presented in Algorithm \ref{AlgSVTD}.

\begin{algorithm}[t]
\caption{Algorithmic code for the time-dependent module.} \label{AlgSVTD}
\begin{center}
\begin{boxedminipage}{13cm}
\begin{algorithmic}
    \STATE Initialize time: $t=t_o$
    \STATE Initialize $\mathbf{U}^{t_o}$
    \FOR {$j \leftarrow 1$ to $numberOfTimeSteps$}
        \STATE Increment time by $\Delta t$
        \FOR {$i \leftarrow 1$ to $MaximumNumberOfNewtonIterations$}
            \STATE Find $\mathbf{R}_{TD}^{t+\Delta t}=\int_{\Omega}\frac{\mathbf{U}^{t+\Delta t}-\mathbf{U}^{t}}{\Delta t}\cdot \mathbf{\omega} dx+\mathbf{R}_{TI}^{t+\Delta t}=\mathbf{0}$
            \STATE Find $\mathbf{J}_{TD}^{t+\Delta t}=\frac{\partial\mathbf{R}_{TD}^{t+\Delta t}}{\partial\mathbf{U}^{t+\Delta t}}$
            \STATE Find $\Delta\mathbf{U}^{t+\Delta t}=-\left(\mathbf{J}^{-1}\mathbf{R}\right)_{TD}^{t+\Delta t}$
            \STATE Find $\mathbf{U}_{i+1}^{t+\Delta t}=\mathbf{U}_{i}^{t+\Delta t}+\Delta\mathbf{U}^{t+\Delta t}$
            \STATE Update: $\mathbf{U}_{i}^{t+\Delta t}=\mathbf{U}_{i+1}^{t+\Delta t}$
            \IF{(convergence condition met)}
                \STATE {Exit loop}
            \ELSE
                \IF{($MaximumNumberOfNewtonIterations$ reached)}
                    \STATE {Declare failure}
                    \STATE {Exit}
                \ENDIF
            \ENDIF
        \ENDFOR
        \STATE Update: $\mathbf{U}^{t}=\mathbf{U}^{t+\Delta t}$
    \ENDFOR
    \STATE Solution: $\mathbf{U}^{t+\Delta t}$
    \STATE End
\end{algorithmic}
\end{boxedminipage}
\end{center}
 \vskip17.5pt
\end{algorithm}

\subsection{Branched Network}

To extend the time-independent and time-dependent single vessel model to time-independent and
time-dependent branched network of interconnected vessels, matching constraints at the branching
nodes are required. These nodes are treated as discontinuous joints where each segment connected to
that junction has its own index for that junction although they are spatially identical. The
matching constraints are derived from the conservation of flow rate for incompressible fluid, and
the Bernoulli energy conservation principle for inviscid flow. More specifically, at each
$n$-segment branching node, $n$ distinctive constraints are imposed: one represents the
conservation of flow which involves all the segments at that junction, while the other ($n-1$)
constraints represent the Bernoulli principle with each Bernoulli constraint involving two
distinctive segments. These constrains are summarized in the following relations
\begin{equation}\label{ConQ}
\sum_{i=1}^{n}Q_{i}=0
\end{equation}
and

\begin{equation}\label{Bernoulli}
p_{k}+\frac{1}{2}\rho u_{k}^{2}-p_{l}-\frac{1}{2}\rho u_{l}^{2}=0
\end{equation}
where $k$ and $l$ are indices of two distinct segments, and $u$ ($=\frac{Q}{A}$) is the fluid speed
averaged over the vessel cross section. In Equation \ref{ConQ} a directional flow is assumed by
attaching opposite signs to the inflow and outflow. The matching constraints, which replace the
residuals of one of the governing equations (continuity), are coupled with compatibility
conditions, similar to the ones used for the single vessel, where these conditions replace the
residual of the other governing equation (momentum). The sign convention for these compatibility
conditions should follow the same rules as for the boundary conditions, that is minus sign for
inflow and plus sign for outflow. This branching model can be applied to any branching node with
connectivity $n\geq2$. The special case of $n=2$ enables flexible modeling of discontinuous
transition between two neighboring segments with different cross sectional areas. Suitable pressure
or flux boundary conditions (which for the time-dependent case could be time-independent, or
time-dependent over the whole or part of the time stepping process) should also be imposed on all
boundary nodes of the network. With regard to the other aspects of the time-independence and
time-dependence treatment, the network model should follow the same rules as for single vessel
time-independent and time-dependent models which are outlined in the previous sections.

\section{Non-Dimensionalized Form}\label{NonDimensionalization}

To improve convergence, the aforementioned dimensional forms of the governing, boundary,
compatibility, and matching equations can be non-dimensionalized by carefully-chosen scale factors.
The following scale factors are commonly used to scale the model parameters:

\begin{equation}\label{scaleFacEq}
Q\sim\pi R_{o}^{2}U_{o}\,\,\,\,\,\,\,\,\,\, A\sim\pi R_{o}^{2}\,\,\,\,\,\,\,\,\,\, p\sim\rho
U_{o}^{2}\,\,\,\,\,\,\,\,\,\, x\sim\lambda\,\,\,\,\,\,\,\,\,\, t\sim\frac{\lambda}{U_{o}}
\end{equation}
where $R_o$, $U_o$, and $\lambda$ are respectively typical values of the radius, velocity and
length for the flow system. In the following we demonstrate non-dimensionalization of the flow
equations by a few examples followed by stating the non-dimensionalized form of the others.

\subsection{Non-Dimensionalized Navier-Stokes Equations}

\vspace{0.5cm} \noindent \underline{Continuity equation 1st form (Equation \ref{ConEq1}):}

\begin{equation}
\frac{\partial A}{\partial t}+\frac{\partial Q}{\partial x}=0
\end{equation}

\begin{equation}
\frac{\partial\left(\pi
R_{o}^{2}A'\right)}{\partial\left(\frac{\lambda}{U_{o}}t'\right)}+\frac{\partial\left(\pi
R_{o}^{2}U_{o}Q'\right)}{\partial\left(\lambda x'\right)}=0
\end{equation}
that is

\begin{equation}\label{NDConF1}
\frac{\partial A'}{\partial t'}+\frac{\partial Q'}{\partial x'}=0
\end{equation}
where the prime indicates a non-dimensionalized value.

\vspace{0.5cm} \noindent \underline{Continuity equation 2nd form (Equation \ref{MatrixEq1}):}

Same as Equation \ref{NDConF1}.

\vspace{0.5cm} \noindent \underline{Momentum equation 1st form (Equation \ref{MomEq1}):}

\begin{equation}
\frac{\partial Q}{\partial t}+\frac{\partial}{\partial x}\left(\frac{\alpha
Q^{2}}{A}\right)+\frac{A}{\rho}\frac{\partial p}{\partial x}+\kappa\frac{Q}{A}=0
\end{equation}

\begin{equation}
\frac{\partial\left(\pi
R_{o}^{2}U_{o}Q'\right)}{\partial\left(\frac{\lambda}{U_{o}}t'\right)}+\frac{\partial}{\partial\left(\lambda
x'\right)}\left(\frac{\alpha\left(\pi R_{o}^{2}U_{o}Q'\right)^{2}}{\left(\pi
R_{o}^{2}A'\right)}\right)+\frac{\left(\pi R_{o}^{2}A'\right)}{\rho}\frac{\partial\left(\rho
U_{o}^{2}p'\right)}{\partial\left(\lambda x'\right)}+\kappa\frac{\left(\pi
R_{o}^{2}U_{o}Q'\right)}{\left(\pi R_{o}^{2}A'\right)}=0
\end{equation}

\begin{equation}
\frac{\pi R_{o}^{2}U_{o}^{2}}{\lambda}\frac{\partial Q'}{\partial
t'}+\frac{\alpha\pi^{2}R_{o}^{4}U_{o}^{2}}{\pi R_{o}^{2}\lambda}\frac{\partial}{\partial
x'}\left(\frac{Q'^{2}}{A'}\right)+\frac{\pi R_{o}^{2}\rho U_{o}^{2}}{\lambda\rho}\frac{A'\partial
p'}{\partial x'}+\frac{\kappa\pi R_{o}^{2}U_{o}}{\pi R_{o}^{2}}\frac{Q'}{A'}=0
\end{equation}

\begin{equation}
\frac{\pi R_{o}^{2}U_{o}^{2}}{\lambda}\frac{\partial Q'}{\partial t'}+\frac{\alpha\pi
R_{o}^{2}U_{o}^{2}}{\lambda}\frac{\partial}{\partial x'}\left(\frac{Q'^{2}}{A'}\right)+\frac{\pi
R_{o}^{2}U_{o}^{2}}{\lambda}\frac{A'\partial p'}{\partial x'}+\frac{\kappa\lambda\pi
R_{o}^{2}U_{o}^{2}}{\lambda\pi R_{o}^{2}U_{o}}\frac{Q'}{A'}=0
\end{equation}

\begin{equation}
\frac{\partial Q'}{\partial t'}+\alpha\frac{\partial}{\partial
x'}\left(\frac{Q'^{2}}{A'}\right)+\frac{A'\partial p'}{\partial x'}+\frac{\kappa\lambda}{\pi
R_{o}^{2}U_{o}}\frac{Q'}{A'}=0
\end{equation}


\begin{equation}
\frac{\partial Q'}{\partial t'}+\alpha\frac{\partial}{\partial
x'}\left(\frac{Q'^{2}}{A'}\right)+\frac{A'\partial p'}{\partial
x'}+\frac{2\pi\alpha\nu\lambda}{\left(\alpha-1\right)\pi R_{o}^{2}U_{o}}\frac{Q'}{A'}=0
\end{equation}
that is

\begin{equation}
\frac{\partial Q'}{\partial t'}+\alpha\frac{\partial}{\partial
x'}\left(\frac{Q'^{2}}{A'}\right)+\frac{A'\partial p'}{\partial
x'}+\frac{2\alpha\nu\lambda}{\left(\alpha-1\right)R_{o}^{2}U_{o}}\frac{Q'}{A'}=0
\end{equation}

\vspace{0.5cm} \noindent \underline{Momentum equation 2nd form (Equation \ref{MatrixEq1}):}

\begin{equation}
\frac{\partial Q}{\partial t}+\frac{\partial}{\partial x}\left(\frac{\alpha
Q^{2}}{A}\right)+\frac{\beta}{3\rho A_{o}}\frac{\partial}{\partial
x}A^{3/2}+\kappa\frac{Q}{A}=0
\end{equation}

\begin{equation}
\frac{\pi R_{o}^{2}U_{o}^{2}\partial Q'}{\lambda\partial t'}+\frac{\left(\pi
R_{o}^{2}U_{o}\right)^{2}\partial}{\lambda\pi R_{o}^{2}\partial x'}\left(\frac{\alpha
Q'^{2}}{A'}\right)+\frac{\left(\pi R_{o}^{2}\right)^{1/2}\beta}{\lambda3\rho
A'_{o}}\frac{\partial}{\partial x'}A'^{3/2}+\kappa\frac{\pi R_{o}^{2}U_{o}Q'}{\pi
R_{o}^{2}A'}=0
\end{equation}

\begin{equation}
\frac{\pi R_{o}^{2}U_{o}^{2}\partial Q'}{\lambda\partial t'}+\frac{\pi
R_{o}^{2}U_{o}^{2}\partial}{\lambda\partial x'}\left(\frac{\alpha
Q'^{2}}{A'}\right)+\frac{\left(\pi R_{o}^{2}\right)^{1/2}\beta}{\lambda3\rho
A'_{o}}\frac{\partial}{\partial x'}A'^{3/2}+\kappa\frac{U_{o}Q'}{A'}=0
\end{equation}

\begin{equation}
\frac{\partial Q'}{\partial t'}+\frac{\partial}{\partial x'}\left(\frac{\alpha
Q'^{2}}{A'}\right)+\frac{1}{\left(\pi R_{o}^{2}\right)^{1/2}U_{o}^{2}}\frac{\beta}{3\rho
A'_{o}}\frac{\partial}{\partial x'}A'^{3/2}+\frac{\lambda}{\pi
R_{o}^{2}U_{o}}\kappa\frac{Q'}{A'}=0
\end{equation}

\subsection{Non-Dimensionalized Compatibility Condition}

\vspace{0.5cm} \noindent \underline{Time-independent term of compatibility condition:}

{\scriptsize

\begin{eqnarray}\label{NDComTIF1}
\left(-\alpha\frac{U_{o}Q'}{A'}\pm\sqrt{\frac{U_{o}^{2}Q'^{2}}{A'^{2}}\left(\alpha^{2}-\alpha\right)+\frac{A'}{\rho\sqrt{\pi R_{o}^{2}}}\frac{\beta}{2A'_{o}\sqrt{A'}}}\right)\frac{U_{o}\partial Q'}{\lambda\partial x'}\nonumber\\
+\left(-\alpha\frac{U_{o}^{2}Q'^{2}}{A'^{2}}+\frac{\beta}{2\rho\sqrt{\pi
R_{o}^{2}}A'_{o}}A'^{1/2}\right)\frac{\partial A'}{\lambda\partial x'}+\left(2\alpha\frac{\pi
R_{o}^{2}U_{o}\partial Q'}{\lambda\partial
x'}+\frac{2\pi\alpha\nu}{\alpha-1}\right)\frac{U_{o}Q'}{\pi R_{o}^{2}A'} & = & 0
\end{eqnarray}
}

\vspace{0.5cm} \noindent \underline{Time-dependent term of compatibility condition:}

{\scriptsize
\begin{equation}\label{NDComTDF1}
\left(-\alpha\frac{U_{o}Q'}{A'}\pm\sqrt{\frac{U_{o}^{2}Q'^{2}}{A'^{2}}\left(\alpha^{2}-\alpha\right)+\frac{A'}{\rho}\frac{\beta}{\left(\pi
R_{o}^{2}\right)^{1/2}2A'_{o}\sqrt{A'}}}\right)\frac{\partial A'}{\partial t'}+\frac{U_{o}\partial
Q'}{\partial t'}=0
\end{equation}
}

\subsection{Non-Dimensionalized Matching Conditions}

\noindent \underline{Flow conservation:}

\begin{equation}
Q'_{1}-Q'_{2}-Q'_{3}=0
\end{equation}

\vspace{0.5cm} \noindent \underline{Bernoulli:}

\begin{equation}
p'_{k}+\frac{1}{2}{u'}_{k}^{2}-p'_{l}-\frac{1}{2}{u'}_{l}^{2}=0
\end{equation}

\subsection{Non-Dimensionalized Boundary Conditions}

\begin{equation}
A'-A'_{BC}=0 \hspace{0.5cm} ({\rm for\,\,area\,\,BC}) \hspace{0.5cm} \& \hspace{0.5cm} Q'-Q'_{BC}=0
\hspace{0.5cm} ({\rm for\,\,flow\,\,BC})
\end{equation}

\section{Validation}\label{Validation}

The different modules of the the 1D finite element flow model are validated as follow

\begin{itemize}

\item
Time-independent single vessel: the numeric solution should match the analytic solution as given by
Equation \ref{AnalEq} which is derived in Appendix A. Also, the boundary conditions should be
strictly satisfied.

\item
Time-dependent single vessel: the solution should asymptotically converge to the analytic solution
on imposing time-independent boundary conditions. Also, the boundary conditions should be strictly
satisfied at all time steps.

\item
Time-independent network: four tests are used to validate the numeric solution. First, the boundary
conditions should be strictly satisfied. Second, the conservation of mass (or conservation of
volume for incompressible flow), as given by Equation \ref{ConQ}, should be satisfied at all
branching nodes (bridge, bifurcation, trifurcation, etc.). A consequence of this condition is that
the sum of the boundary inflow (sum of $Q$ at inlet boundaries) should be equal to the sum of the
boundary outflow (sum of $Q$ at outlet boundaries). Third, the conservation of energy (Bernoulli's
principle for inviscid flow), as given by Equation \ref{Bernoulli}, should be satisfied at all
branching nodes. Fourth, the analytic solution for time-independent flow in a single vessel, as
given by Equation \ref{AnalEq} in Appendix A, should be satisfied by all vessels in the network
with possible exception of very few vessels with odd features (e.g. those with distorted shape such
as extreme radius-to-length ratio, and hence are susceptible to large numerical errors). The fourth
test is based on the fact that the single vessel solution is dependent on the boundary conditions
and not on the mechanism by which these conditions are imposed.

\item
Time-dependent network: the solution is validated by asymptotic convergence to the time-independent
solution, as validated by the four tests outlined in the previous item, on imposing
time-independent boundary conditions.

\end{itemize}

The solutions may also be tested qualitatively by static and dynamic visualization for
time-independent and time-dependent cases respectively to verify their physical sensibility. Other
qualitative tests, such as comparing the solutions of different cases with common features, may
also be used for validation.

It should be remarked that Equation \ref{AnalEq} contains three variables: $x$ $A$ and $Q$, and
hence it can be solved for one of these variables given the other two. Solving for $A$ and $Q$
requires employing a numeric solver, based for example on a bisection method; hence the best option
is to solve for $x$ and compare to the numeric solution. This in essence is an exchange of the role
of independent and dependent variables which has no effect on validation. Alternatively, Equation
\ref{QElastic} can be used to verify the solution directly by using the vessel inlet and outlet
areas. In fact Equation \ref{QElastic} can be used to verify the solution at any point on the
vessel axis by labeling the area at that point as $A_{ou}$, as explained in Appendix A.

\section{General Notes}

\subsection{Implementation}\label{Implementation}

\begin{itemize}

\item
The model described in this report was implemented and tested on both single vessels and networks
of vessels for time-independent and time-dependent cases and it produced valid results. The
implementation is based on a Galerkin method, where the test functions are obtained from the same
space as the trial basis functions used to represent the state variables, with a Lagrange
interpolation associated with a Gauss quadrature integration scheme (refer to Appendix B for Gauss
quadrature tables). Many tests have been carried out to verify various aspects of the 1D model.
These tests involved many synthetic and biological networks which vary in their size, connectivity,
number and type of branching nodes, type of meshing, and so on. The tests also included networks
with and without loops although the great majority of the networks contain loops. Some of the
networks involved in these tests consist of very large number of vessels in the order of hundreds
of thousands with much more degrees of freedom. Non-dimensional form, as well as dimensional form,
was tested on single vessels and branched networks; the results, after re-scaling, were verified to
be identical to those obtained from the dimensional form. The checks also included $h$ and $p$
convergence tests which demonstrated correct convergence behavior.

\item
To be on the safe side, the order of the quadrature should be based on the sum of orders of the
interpolating functions, their derivatives and test functions. The adopted quadrature order scheme
takes the highest order required by the terms.

\item
A constant delta may be used in the evaluation of the Jacobian matrix by finite difference. A
suitable value for delta may be $\Delta A_i = \Delta Q_i = 10^{-7}$ or $10^{-8}$.

\end{itemize}

\subsection{Solution}

\begin{itemize}

\item
Negative flow in the solution means the flow direction is opposite to the vessel direction as
indicated by the vessel topology, that is the flow rate of a segment indexed as $N_{1}$ $N_{2}$
will be positive if the flow is from $N_{1}$ to $N_{2}$ and negative if the flow is from $N_{2}$ to
$N_{1}$.

\item
Interpolation polynomials of various degrees ($p$) in association with different meshing ($h$)
should be used to validate the convergence behavior. The convergence to the correct solution should
improve by increasing $p$ and decreasing $h$. $L^2$ error norm may be used as a measure for
convergence; it is given by

\begin{equation}
L^{2}=\left(\int_{X}\left(S_{a}-S_{n}\right)^{2}dx\right)^{1/2}
\end{equation}
where $S_{a}$ and $S_{n}$ are the analytic and numeric solutions respectively, and $X$ is the
solution domain. The integration can be performed numerically using, for example, trapezium or
Simpson's rules. The error norm should fall steadily as $h$ decreases and $p$ increases.

\item
With regard to the previously outlined implementation of the 1D model (see \S\
\ref{Implementation}), typical solution time on a typical platform (normal laptop or desktop) for a
single time-independent simulation on a typical 1D network consisting of hundreds of thousands of
degrees of freedom is a few minutes. The final convergence is normally reached within 5-7 Newton
iterations. The solution time of a single time step for the time-dependent case is normally less
than the solution time of the equivalent time-independent case, and the number of the required
Newton iterations of each time step in the time-dependent case is normally less than that for the
corresponding time-independent case.

\item
Since there are many sources of error and wrong convergence, each acquired solution should be
verified by the aforementioned validation tests (see \S\ \ref{Validation}). The 1D finite element
code should be treated as a device that suggests solutions which can be accepted only if they meet
the validation criteria.

\end{itemize}

\subsection{Non-Dimensionalization}

\begin{itemize}

\item
On implementing the non-dimensionalized form (as given in \S\ \ref{NonDimensionalization}) in the
finite element code, all the user needs is to scale the primed input values either inside or
outside the code; the results then should be scaled back to obtain the dimensionalized solution.

\item
Different length scales can be utilized as long as they are in different orientations (e.g. vessel
length and vessel radius) and hence linearly independent; otherwise the physical space will be
distorted in non-systematic way and hence may not be possible to restore by scaling back.

\end{itemize}

\subsection{Convergence}

A number of measures, outlined in the following points, can speedup convergence and help avoiding
convergence failure.

\begin{itemize}

\item
Non-dimensionalization which requires implementation in the finite element code (as given in \S\
\ref{NonDimensionalization}) where the input data is non-dimensionalized and the results are
re-dimensionalized back to the physical space.

\item
Using different unit systems, such as m.kg.s or mm.g.s or m.g.s, for the input data and parameters.

\item
Scaling the model up or down to obtain a similarity solution which can then be scaled back to
obtain the final results.

\item
Increasing the error tolerance of the solver for convergence criterion. However, the use of
relatively large error tolerance can cause wrong convergence and hence should be avoided. It may be
recommended that the maximum allowed error tolerance for obtaining a reliable solution must not
exceed $10^{-5}$. Anyway, the solution in all cases should be verified by the validation tests (see
\S\ \ref{Validation}) and hence it must be rejected if the errors exceed acceptable limits.

\item
For time-dependent cases, the required boundary condition value can be imposed gradually by
increasing the inlet pressure, for instance, over a number of time steps to reach the final steady
state value.

\item
The use of smaller time steps in the time-dependent cases may also help to avoid convergence
failure.

\end{itemize}

It should be remarked that the first three strategies are based on the same principle, that is
adjusting the size of the problem numbers to help the solver to converge more easily to the
solution.

\subsection{Boundary Conditions}

\begin{itemize}

\item
Dirichlet type boundary conditions are usually used for imposing flow rate and pressure boundary
conditions. The previous formulation is based on this assumption.

\item
Pressure boundary conditions are imposed by adjusting the inlet or outlet area where $p$ and $A$
are correlated through Equation \ref{PAEq}.

\item
While pressure boundary conditions can be imposed on both inlet and outlet boundaries
simultaneously, as well as mixed boundary conditions (i.e. inlet pressure with outlet flow or inlet
flow with outlet pressure or mixed on one or both boundaries), it is not possible to impose flow
boundary conditions on all inlet and outlet boundaries simultaneously because this is {\it either}
a trivial condition repeating the condition of flow conservation (i.e. Equation \ref{ConQ}) at the
branching junctions if the inflow is equal to the outflow {\it or} it is a contradiction to the
flow conservation condition if the inflow and outflow are different, and hence no solution can be
found due to ambiguity and lack of constraints in the first case and to inconsistency in the second
case.

\item
Zero $Q$ boundary condition can be used to block certain inlet or outlet vessels in a network for
the purpose of emulating a physical situation or improving convergence when the blockage does not
affect the solution significantly.

\item
In some biological flow conditions there are no sufficient data to impose realistic pressure
boundary conditions that ensure biologically sensible flow in the correct direction over the whole
vascular network. In such situations a back flow may occur in some branches which is physically
correct but biologically incorrect. To avoid this situation, an inlet pressure boundary condition
with outlet flow boundary conditions where the total outflow is split according to a certain
physical or biological model (such as being proportional to the area squared) can be used to ensure
sensible flow in the right direction over the whole network. The total amount of the outflow can be
estimated from the inlet flow which is usually easier to estimate as it normally comes from a
single (or few) large vessel. This trick may also be applicable in some physical circumstances.

\item
Use may be made of an artificial single inlet boundary to avoid lack of knowledge about the
pressure distribution in a multi-inlet network to ensure correct flow in the right direction. The
inlets can be connected to a single artificial node (e.g. located at their centroid) where the
radii of the connecting artificial vessels is chosen according to a physical or biological model
such as Murray's law. This node can then be connected through a single artificial vessel whose
radius can be computed from a physical or biological model and whose length can be determined from
a typical $L/R$ ratio such as 10. The inlet of this vessel can then be used to impose a single $p$
or $Q$ biologically-sensible boundary condition. It should be remarked that Murray's biological law
is given by

\begin{equation}
r_{m}^{\gamma}=\sum_{i}^{n}r_{d_{i}}^{\gamma}
\end{equation}
where $r_{m}$ is the radius of the mother vessel, $r_{d_{i}}$ is the radius of the $i$th daughter
vessel, $n$ is the number of daughter vessels which in most cases is 2, and $\gamma$ is a constant
index which according to Murray is 3, but other values like 2.1 and 2.2 are also used in the
literature.

\item
Time-dependent boundary conditions can be modeled by empirical signals (e.g. obtained from
experimental data) or by closed analytical forms such as sinusoidal.

\end{itemize}

\subsection{Initial Conditions}

\begin{itemize}

\item
The convergence usually depends on the initial values of area and flow rate. A good option for
these values is to use unstressed area with zero flow for start.

\end{itemize}

\subsection{Miscellaneous}

\begin{itemize}

\item
Apart from the interpolation nodes, there are two main types of nodes in the finite element
network: segment nodes and finite element discretization nodes. The connectivity of the second type
is always 2 as these nodes connect two elements; whereas the connectivity of the first can be 1 for
the boundary nodes, 2 for the bridge nodes connecting two segments, or $\ge3$ for the branching
nodes (bifurcation, trifurcation, etc.). The mass and energy conservation conditions can be
extended to include all the segment nodes with connectivity $>1$ by including the bridge nodes.

\item
For networks, the vessel wall thickness at reference pressure, $h_o$, can be a constant or vary
from vessel to vessel depending on the physical or biological situation. Using variable thickness
is more sound in biological context where the thickness can be estimated as a fraction of the lumen
or vessel radius. A fractional thickness of 10-15\% of the radius is commonly used for blood
vessels \cite{PodesserNNSWM1998, FormaggiaGNQ2001, ZhangFG2002, BlondelLLDBS2003, FormaggiaLQ2003,
WaiteFBook2007, LeeS2008, BadiaQQ2009, BroekHRV2011}. For more details, refer to Appendix D.

\item
The previous finite element formulation of the 1D model for single vessels and networks works for
constant-radius vessels (i.e. with constant $A_o$) only and hence to extend the formulation to
variable-radius vessels the previous matrix structure should be reshaped to include the effect of
tapering or expanding of the vessels. However, the vessels can be straight or curved. The size of
the vessels in a network can also vary significantly from one vessel to another as long as the 1D
flow model assumptions (e.g. size, shape, etc.) do apply on each vessel.

\item
The networks used in the 1D flow model should be totally connected, that is any node in the network
can be reached from any other node by moving entirely inside the network vessels.

\item
Different time stepping schemes, such as forward or backward Euler or central difference, can be
used for implementing the time term of the time-dependent single vessel and network modules
although the speed of convergence and quality of solution vary between these schemes. The size of
the time step should be chosen properly for each scheme to obtain equivalent results from these
different schemes.

\item
Although the 1D model works on highly non-homogeneous networks in terms of vessels length without
discretization, a scheme of homogeneous discretization may be employed by using a constant element
length, $h$, over the whole network as an approximation to the length of the discretized elements.
The length of the elements of each vessel is then obtained by dividing the vessel evenly to an
integer number of elements with closest size to the given $h$. Although discretization is not a
requirement, since the 1D model works even on non-discretized networks, it usually improves the
solution. Moreover, discretization is required for obtaining a detailed picture of the pressure and
flow fields at the interior points. Use of interpolation schemes higher than linear (with and
without discretization) also helps in refining the variable fields. Also, for single vessel the
solution can be obtained with and without discretization; in the first case the discretized
elements could be of equal or varying length. The solution, however, should generally improve by
discretization.

\item
Although the 1D model works on non-homogeneous networks in terms of vessels radius, an abrupt
transition from one vessel to its neighbor may hinder convergence.

\item
In general, the time-dependent problem converges more easily than its equivalent time-independent
problem. This may be exploited to obtain approximate time-independent solutions in some
circumstances from the time-dependent module as the latter asymptotically approaches the
time-independent solution.

\item
The correctness of the solutions mathematically may not guarantee physiological, and even physical,
sensibility since the network features, boundary conditions, and model parameters which in general
highly affect the flow pattern, may not be found normally in real biological and physical systems.
The quality of any solution, assuming its correctness in mathematical terms, depends on the quality
of the underlying model and how it reflects the physical reality.

\item
Because the 1D model depends on the length of the vessels but not their location or orientation, a
1D coordinate system, as well as 2D or 3D, can be used for coordinating the space. The vessels can
be randomly oriented without effecting the solution. A multi-dimensional space may be required,
however, for consistent and physically-correct description of the networks.

\item
The reference pressure, $p_o$, in Equation \ref{PAEq} is usually assumed zero to simplify the
relation.

\end{itemize}

\newpage
\section{Conclusions}

The one-dimensional Navier-Stokes formulation is widely used as a realistic model for the flow of
Newtonian fluids in large vessels with certain simplifying assumptions, such as axi-symmetry. The
model may also be coupled with a pressure-area constitutive relation and hence be extended to the
flow in distensible vessels. Numerical implementation of this model based on a finite element
method with suitable boundary conditions is also used to solve the time-independent and transient
flow in single vessels and networks of interconnected vessels where in the second case
compatibility and matching conditions, which include conservation of mass and energy, at branching
nodes are imposed. Despite its comparative simplicity, the 1D flow model can provide reliable
solutions, with relatively low computational cost, to many flow problems within its domain of
validity. The current document outlined the analytical and numerical aspects of this model with
theoretical and technical details related to implementation, performance, methods of improvement,
validation, and so on.

\clearpage
\phantomsection \addcontentsline{toc}{section}{Nomenclature} %
{\noindent \LARGE \bf Nomenclature} \vspace{0.5cm}

\begin{supertabular}{ll}
$\alpha$                &   correction factor for axial momentum flux \\
$\beta$                 &   parameter in the pressure-area relation \\
$\gamma$                &   Murray's law index \\
$\epsilon$              &   residual error \\
$\zeta$                 &   quadrature point coordinate \\
$\kappa$                &   viscosity friction coefficient \\
$\mu$                   &   fluid dynamic viscosity \\
$\nu$                   &   fluid kinematic viscosity ($\nu = \frac{\mu}{\rho}$) \\
$\rho$                  &   fluid mass density \\
$\varsigma$             &   Poisson's ratio of vessel wall \\
$\psi$                  &   basis function for finite element discretization \\
$\boldsymbol{\omega}$   &   vector of test functions in the weak form of finite element \\
$\Omega$                &   solution domain \\
$\partial \Omega$       &   boundary of the solution domain \\
\\
$A$                     &   vessel cross sectional area \\
$A_{BC}$                &   boundary condition for vessel cross sectional area \\
$A_{in}$                &   vessel cross sectional area at inlet \\
$A_o$                   &   vessel cross sectional area at reference pressure \\
$\mathbf{B}$            &   matrix of force terms in the 1D Navier-Stokes equations \\
$E$                     &   Young's modulus of vessel wall \\
$f(A)$                  &   function in pressure-area relation \\
$\mathbf{F}$            &   matrix of flux quantities in the 1D Navier-Stokes equations \\
$h$                     &   length of element \\
$h_o$                   &   vessel wall thickness at reference pressure \\
$\mathbf{H}$            &   matrix of partial derivative of $\mathbf{F}$ with respect to $\mathbf{U}$ \\
$\mathbf{J}$            &   Jacobian matrix \\
$L$                     &   length of vessel \\
$\mathfrak{N}$          &   norm of residual vector \\
$p$                     &   local pressure \\
$p$                     &   order of interpolating polynomial \\
$p_o$                   &   reference pressure \\
$q$                     &   dummy index for quadrature point \\
$Q$                     &   volumetric flow rate \\
$Q_{BC}$                &   boundary condition for volumetric flow rate \\
$r$                     &   radius \\
$\mathbf{R}$            &   weak form of residual vector \\
$S_{a}$                 &   analytic solution \\
$S_{n}$                 &   numeric solution \\
$t$                     &   time \\
$\Delta t$              &   time step \\
$u$                     &   local axial speed of fluid \\
$\overline{u}$          &   mean axial speed of fluid \\
$\mathbf{U}$            &   vector of finite element variables \\
$\mathbf{\Delta U}$     &   vector of change in $\mathbf{U}$ \\
$x$                     &   vessel axial coordinate \\

\end{supertabular}

\newpage
\phantomsection \addcontentsline{toc}{section}{References} %
\bibliographystyle{unsrt}

\newpage
\section[Appendix A: Derivation of Analytical Solution]{Appendix A: Derivation of Time-Independent Analytical Solution for Single Vessel}

The following analytical relation linking vessel axial coordinate $x$ to cross sectional area $A$,
cross sectional area at inlet $A_{in}$, and volumetric flow rate $Q$ for time-independent flow can
be derived and used to verify the finite element solution

\begin{equation}\label{AnalEq}
x=\frac{\alpha Q^{2}\ln\left(A/A_{in}\right)-\frac{\beta}{5\rho
A_{o}}\left(A^{5/2}-A_{in}^{5/2}\right)}{\kappa Q}
\end{equation}

The derivation is outlined in the following. For time-independent flow, the system given by
Equation \ref{MatrixEq1} in matrix form, will become

\begin{equation}
\frac{\partial Q}{\partial x}=0\,\,\,\,\,\,\,\,\,\,\,\,\, x\in\left[0,l\right],\,\,\,
t\ge0
\end{equation}

\begin{equation}
\frac{\partial}{\partial x}\left(\frac{\alpha Q^{2}}{A}+\frac{\beta}{3\rho
A_{o}}A^{3/2}\right)+\kappa\frac{Q}{A}=0\,\,\,\,\,\,\,\,\,\,\,\,\, x\in\left[0,l\right],\,\,\,
t\ge0
\end{equation}
that is $Q$ as a function of $x$ is constant and

\begin{equation}
\frac{\partial}{\partial A}\left(\frac{\alpha Q^{2}}{A}+\frac{\beta}{3\rho
A_{o}}A^{3/2}\right)\frac{\partial A}{\partial x}+\kappa\frac{Q}{A}=0
\end{equation}
i.e.

\begin{equation}
\left(-\frac{\alpha Q^{2}}{A^{2}}+\frac{\beta}{2\rho A_{o}}A^{1/2}\right)\frac{\partial A}{\partial
x}+\kappa\frac{Q}{A}=0
\end{equation}
which by algebraic manipulation can be transformed to

\begin{equation}
\frac{\partial x}{\partial A}=\frac{-\frac{\alpha Q^{2}}{A}+\frac{\beta}{2\rho
A_{o}}A^{3/2}}{-\kappa Q}
\end{equation}

On integrating the last equation we obtain

\begin{equation}
x=\frac{\alpha Q^{2}\ln A-\frac{\beta}{5\rho A_{o}}A^{5/2}}{\kappa Q}+C
\end{equation}
where $C$ is the constant of integration which can be determined from the boundary condition at
$x=0$ with $A=A_{in}$, that is

\begin{equation}
C=\frac{-\alpha Q^{2}\ln A_{in}+\frac{\beta}{5\rho A_{o}}A_{in}^{5/2}}{\kappa Q}
\end{equation}
i.e.

\begin{equation}
x=\frac{\alpha Q^{2}\ln\left(A/A_{in}\right)-\frac{\beta}{5\rho
A_{o}}\left(A^{5/2}-A_{in}^{5/2}\right)}{\kappa Q}
\end{equation}

For practical reasons, this relation can be re-shaped and simplify to reduce the number of
variables by the use of the second boundary condition at the outlet, as outlined in the following.
When $x=L$, $A=A_{ou}$ where $L$ is the vessel length and $A_{ou}$ is the cross sectional area at
the outlet, that is

\begin{equation}
L=\frac{\alpha Q^{2}\ln\left(A_{ou}/A_{in}\right)-\frac{\beta}{5\rho
A_{o}}\left(A_{ou}^{5/2}-A_{in}^{5/2}\right)}{\kappa Q}
\end{equation}
which is a quadratic polynomial in $Q$ i.e.

\begin{equation}
-\alpha\ln\left(A_{ou}/A_{in}\right)Q^{2}+\kappa LQ+\frac{\beta}{5\rho
A_{o}}\left(A_{ou}^{5/2}-A_{in}^{5/2}\right)=0
\end{equation}

\begin{equation}
\alpha\ln\left(A_{in}/A_{ou}\right)Q^{2}+\kappa LQ+\frac{\beta}{5\rho
A_{o}}\left(A_{ou}^{5/2}-A_{in}^{5/2}\right)=0
\end{equation}
with a solution given by

\begin{equation}
Q=\frac{-\kappa L\pm\sqrt{\kappa^{2}L^{2}-4\alpha\ln\left(A_{in}/A_{ou}\right)\frac{\beta}{5\rho
A_{o}}\left(A_{ou}^{5/2}-A_{in}^{5/2}\right)}}{2\alpha\ln\left(A_{in}/A_{ou}\right)}
\end{equation}
which is necessarily real for $A_{in}\ge A_{ou}$ which can always be satisfied for normal flow
conditions by proper labeling. For a flow physically-consistent in direction with the pressure
gradient, the root with the plus sign should be chosen, i.e.

\begin{equation}\label{QElastic}
Q=\frac{-\kappa L+\sqrt{\kappa^{2}L^{2}-4\alpha\ln\left(A_{in}/A_{ou}\right)\frac{\beta}{5\rho
A_{o}}\left(A_{ou}^{5/2}-A_{in}^{5/2}\right)}}{2\alpha\ln\left(A_{in}/A_{ou}\right)}
\end{equation}

This, in essence, is a relation between flow rate and pressure drop (similar to the
Hagen-Poiseuille law for rigid vessels) although for elastic vessels the flow rate, as given by
Equation \ref{QElastic}, does not depend on the pressure difference (as for rigid vessels) but on
the actual inlet and outlet pressures.

Although Equation \ref{QElastic} may look a special case of Equation \ref{AnalEq} as it involves
only the vessel two end areas, $A_{ou}$ may be assumed to be the area at any point along the vessel
axis, with $L$ being the distance form the vessel inlet to that point, and hence this relation can
be used to verify the finite element solution at any point on the vessel.

\newpage
\section[Appendix B: Gauss Quadrature]{Appendix B: Gauss Quadrature}

In this appendix we list points and weights of Gauss quadrature for polynomials of order 1-10 which
may not be easy to find.

\newpage
\setlength{\textheight}{25.2cm}

\begin{landscape}
\begin{table}
\caption{Gauss quadrature points and weights for polynomial order, $p$, 1-10 assuming a 0-1 master
element. \label{Quadrature}} {\tiny
\begin{tabular}{|p{.2cm}|c@{ }|c@{ }|c@{ }|c@{ }|c@{ }|c@{ }|c@{ }|c@{ }|c@{ }|c|}
\hline
{\bf $p$} &                                                                                             \multicolumn{ 10}{c|}{{\bf Points}} \\
\hline
   {\bf 1} & 0.500000000000000 &                   &                   &                   &                   &                   &                   &                   &                   &                   \\

   {\bf 2} & 0.211324865405188 & 0.788675134594812 &                   &                   &                   &                   &                   &                   &                   &                   \\

   {\bf 3} & 0.112701665379259 & 0.500000000000000 & 0.887298334620741 &                   &                   &                   &                   &                   &                   &                   \\

   {\bf 4} & 0.069431844202974 & 0.330009478207572 & 0.669990521792428 & 0.930568155797026 &                   &                   &                   &                   &                   &                   \\

   {\bf 5} & 0.046910077030669 & 0.230765344947159 & 0.500000000000000 & 0.769234655052841 & 0.953089922969331 &                   &                   &                   &                   &                   \\

   {\bf 6} & 0.033765242898424 & 0.169395306766867 & 0.380690406958402 & 0.619309593041598 & 0.830604693233133 & 0.966234757101576 &                   &                   &                   &                   \\

   {\bf 7} & 0.025446043828621 & 0.129234407200303 & 0.297077424311301 & 0.500000000000000 & 0.702922575688699 & 0.870765592799697 & 0.974553956171380 &                   &                   &                   \\

   {\bf 8} & 0.019855071751232 & 0.101666761293187 & 0.237233795041836 & 0.408282678752175 & 0.591717321247825 & 0.762766204958164 & 0.898333238706813 & 0.980144928248768 &                   &                   \\

   {\bf 9} & 0.015919880246187 & 0.081984446336682 & 0.193314283649705 & 0.337873288298095 & 0.500000000000000 & 0.662126711701905 & 0.806685716350295 & 0.918015553663318 & 0.984080119753813 &                   \\

  {\bf 10} & 0.013046735741414 & 0.067468316655508 & 0.160295215850488 & 0.283302302935377 & 0.425562830509185 & 0.574437169490815 & 0.716697697064623 & 0.839704784149512 & 0.932531683344493 & 0.986953264258586 \\
\hline
    {\bf } &                                                                                            \multicolumn{ 10}{c|}{{\bf Weights}} \\
\hline
   {\bf 1} & 1.000000000000000 &                   &                   &                   &                   &                   &                   &                   &                   &                   \\

   {\bf 2} & 0.500000000000000 & 0.500000000000000 &                   &                   &                   &                   &                   &                   &                   &                   \\

   {\bf 3} & 0.277777777777778 & 0.444444444444444 & 0.277777777777778 &                   &                   &                   &                   &                   &                   &                   \\

   {\bf 4} & 0.173927422568727 & 0.326072577431273 & 0.326072577431273 & 0.173927422568727 &                   &                   &                   &                   &                   &                   \\

   {\bf 5} & 0.118463442528095 & 0.239314335249683 & 0.284444444444444 & 0.239314335249683 & 0.118463442528095 &                   &                   &                   &                   &                   \\

   {\bf 6} & 0.085662246189585 & 0.180380786524069 & 0.233956967286345 & 0.233956967286345 & 0.180380786524069 & 0.085662246189585 &                   &                   &                   &                   \\

   {\bf 7} & 0.064742483084435 & 0.139852695744638 & 0.190915025252559 & 0.208979591836735 & 0.190915025252559 & 0.139852695744638 & 0.064742483084435 &                   &                   &                   \\

   {\bf 8} & 0.050614268145188 & 0.111190517226687 & 0.156853322938943 & 0.181341891689181 & 0.181341891689181 & 0.156853322938943 & 0.111190517226687 & 0.050614268145188 &                   &                   \\

   {\bf 9} & 0.040637194180787 & 0.090324080347429 & 0.130305348201467 & 0.156173538520001 & 0.165119677500630 & 0.156173538520001 & 0.130305348201467 & 0.090324080347429 & 0.040637194180787 &                   \\

  {\bf 10} & 0.033335672154344 & 0.074725674575291 & 0.109543181257991 & 0.134633359654998 & 0.147762112357376 & 0.147762112357376 & 0.134633359654998 & 0.109543181257991 & 0.074725674575291 & 0.033335672154344 \\
\hline
\end{tabular}
}
\end{table}
\end{landscape}

\newpage
\section{Appendix C: Biological Parameters}

In this appendix, we suggest some biologically-realistic values for the 1D flow model parameters in
the context of simulating blood flow in large vessels.

\begin{enumerate}

\item
Blood mass density ($\rho$): 1050~kg.m$^{-3}$ \cite{Avolio1980, FormaggiaGNQ2001, SmithPH2002,
CanicK2003, Smith2004, KoshibaACH2007, LeeS2008, AshikagaCYVOC2008, BadiaQQ2009}.

\item
Blood dynamic viscosity ($\mu$): 0.0035~Pa.s \cite{Avolio1980, FormaggiaGNQ2001, SmithPH2002,
AntigaPhDThesis2002, FormaggiaLQ2003, WesterhofBLS2006, KoshibaACH2007, MouraThesis2007, LeeS2008,
AlastrueyPPS2008, BadiaQQ2009, JanelaMS2010}.

\item
Young's elastic modulus ($E$): 100~kPa \cite{Avolio1980, FormaggiaGNQ2001, ZhangFG2002,
FormaggiaLQ2003, ZhangG2006, FormaggiaMN2006, LeventalGJ2007, MouraThesis2007, LeeS2008,
AlastrueyPPS2008, CaloBBH2008, BadiaQQ2009, HunterALLLe2010, JanelaMS2010}. Also see
\cite{DengTDF1994, BadiaQQ2009} on shear modulus.

\item
Vessel wall thickness ($h_o$): this, preferably, is vessel dependent, i.e. a fraction of the lumen
or vessel radius according to some experimentally-established mathematical relation. The relation
between wall thickness and vessel inner radius is somehow complex and vary depending on the type of
vessel (e.g. artery or capillary). For arteries, the typical ratio of wall thickness to inner
radius is about 0.1-0.15, and this ratio seems to go down in the capillaries and arterioles.
Therefore a typical value of 0.1 seems reasonable \cite{PodesserNNSWM1998, FormaggiaGNQ2001,
ZhangFG2002, BlondelLLDBS2003, FormaggiaLQ2003, WaiteFBook2007, LeeS2008, BadiaQQ2009,
BroekHRV2011}.

\item
Momentum correction factor ($\alpha$): $4/3=1.33$ assuming Newtonian flow. A smaller value, e.g.
1.2, may be used to account for non-Newtonian shear-thinning effects \cite{FormaggiaNQV1999,
SmithPH2002, CanicK2003, FormaggiaLQ2003, SherwinFPP2003, FormaggiaLTV2006, LeeS2008,
PasseriniLFQV2009}.

\item
Time step ($\Delta t$): 1.0-0.1~ms \cite{QuarteroniRV2001, SmithPH2002, FormaggiaLQ2003,
PontrelliR2003, UrquizaBLVF2003, Smith2004, FormaggiaMN2006, MarieCCS2006, BlancoFU2007,
MouraThesis2007, LeeS2008, AlastrueyPPS2008, BadiaQQ2009, BoulakiaCFGZ2010, Papadakis2009,
JanelaMS2010, ChapelleGMC2010, CarloNPT}.

\item
Pressure step ($\Delta p$): 1.0-5.0~kPa \cite{SmithPH2002, Smith2004, LeeS2008, JanelaMS2010}.

\item
Time of heart beat: 0.85~s assuming 70 beats per minute.

\item
Poisson ratio ($\varsigma$): 0.45 \cite{Avolio1980, DengTDF1994, FormaggiaGNQ2001,
QuarteroniRV2001, ZhangFG2002, SherwinFPP2003, FormaggiaMN2006, MouraThesis2007, CaloBBH2008,
BadiaQQ2009, JanelaMS2010}.

\end{enumerate}

\end{document}